# The Virtual Neutron Experiment for TOF Neutron Reflectometer

Taisen Zuo[1,2,3], Haolai Tian [1,2 ;1)], Yanyan Wang [1,2], Rong Du [1,2], Ming Tang[1,2], Junrong Zhang [1,2;2)], Fang-Wei Wang [1,4], Yuan-Bo Chen[1,2]

[1] China Spallation Neutron Source (CSNS), Institute of High Energy Physics (IHEP), Chinese Academy of Sciences (CAS), Dongguan 523803, China

[2] Dongguan Institute of Neutron Science (DINS), Dongguan 523808, China

[3] University of Chinese Academy of Sciences, Beijing 100049, China

[4] Institute of Physics, Chinese Academy of Sciences, Beijing 100190, China

**Abstract:** A general virtual neutron experiment for TOF neutron reflectometer was introduced, including instrument simulation, sample modeling, detector simulation and data reduction to mimic the routine of real experimental process and data reduction. The reduced data fit quite well with the sample simulation, confirming the reliability of our algorithm and the smearing effect was analyzed. Methodology and algorithms developed in this study paves the way for the future development of data processing and data reduction software for the TOF neutron reflectometers in China Spallation Neutron Source (CSNS). Details of the virtual experiment and performance of the instrument was demonstrated.

## 1 Introduction

Since the discovery of neutron by Chard Wick in 1936, people soon realize the optical properties of slow neutrons analogue to the refraction, reflection and interference of light[1]. And soon development of multilayer supermirror[2] followed. Early instrumentation based on supermirror were built in ILL and Saclay. But it was not until after 1980's, dedicated reflectometer were built at ISIS(CRISP) [3], Scalay (PRISM)[4] and IPNS(POSY) [5] that neutron reflectivity start to take the shape we recognize now. Groups who had never used neutron scattering joint the new community of neutron reflectometry. The community consists of two main groups, Soft Matter application group and hard condensed matter (magnetic thin films and multilayers) application group[6]. CRISP, SURF[7], and INTER[8], three generation of Soft Matter dedicated reflectometer in ISIS forged generations of pioneers of neutron reflectometry in UK and the global community such as Jeff Penfold, Rob Richardson, John White, Julia Higgins et al. Hard condensed matter application of neutron reflectivity is more based on the development of theory and technology. Mezei et al. developed polarizing supermirrors[2] which was a key of carrying Polarized Neutron Reflectivity(PNR). Theory for PNR was developed by Mendiratta and Bulume, Sivardiere, Belyakov and Bokun[9] et al. With the development of techniques, polarized neutron will play an vital role in the field of spintronics[10] superconducting and multiferroic materials.

In the last two decades, big neutron centers and brighter neutron sources are being built as versatile platforms for multi-discipline research. Reflectometers are usually day-one instruments and play an indispensable part in the instrument suite to serve the neutron scattering community. For example, the Magnetism Reflectometer (MR) [11] and Liquid Reflectometer (LR) [12] had being built in Spallation Neutron Source (SNS) of the United States in 2006, and four reflectometers had been built

* Supported by National Natural Science Foundation of China (11305191)

1) E-mail:tianhl@ihep.ac.cn

2) E-mail: jrzhang@ ihep.ac.cn

in ISIS in the United Kingdom[13]. In China, two research reactors and one spallation neutron source equipped with advanced neutron instrumentations are being built to join the world's clubs of neutron scattering. Eight neutron instruments had been built in the 60MW Chinese Advanced Research Reactor (CARR) [14, 15], including one liquid reflectometer[16]. China Mianyang Research Reactor also have eight instruments developed including one vertical sample polarized TOF reflectometer[17]. Two reflectometers, one day one magnetism reflectometer with vertical sample geometry [18] and another liquid reflectomter with horizontal sample geometry are estimated to be built in CSNS.

Thanks to the development of the Monte-Carlo simulation packages such as VITESS[19] and McStas[20, 21], optical components and the whole instrument can be simulated and optimized for specific situations [22]. Together with the maturity of the data reduction platform such as Mantid[23], it is now possible to conduct virtual experiments based on these packages with acceptable accuracy and expense[24, 25]. In the United States and Europe, projects have been initiated to support the development of digital infrastructures for virtual experiments[24, 26]. In this work, a whole reflectometry virtual experiment was conducted to mimic a real experiment as close as possible. However, it is complicated or impossible to simulate every detail of a real experiment, like incoherent scattering from the sample, natural background or polarized neutron reflectometry, but the basic principles and process of data reduction should be the same.

The goal of this work is to reproduce the process of a reflectometry experiment by means of simulations, verifying the design of instrument, validating the correctness of data processing software, training the users and preparing for the challenge of the data from real experiments.

## 2 Experimental method

The virtual neutron experiment was done based mainly on mature software packages and in-house software packages DroNE[27] (Data pROcessing suit for Neutron Experiments). Data flow of the virtual experiment was shown in Fig 1, including instrument simulation, detector simulation, data processing and data reduction. Based on the source data of the coupled hydrogen moderator provided by the neutronics group of CSNS [28], a virtual reflectometer was built in VITESS and neutron trajectories were traced from the moderator to the detector. All the neutron trajectories at the detector were written out as input of DroNE for detector simulation.

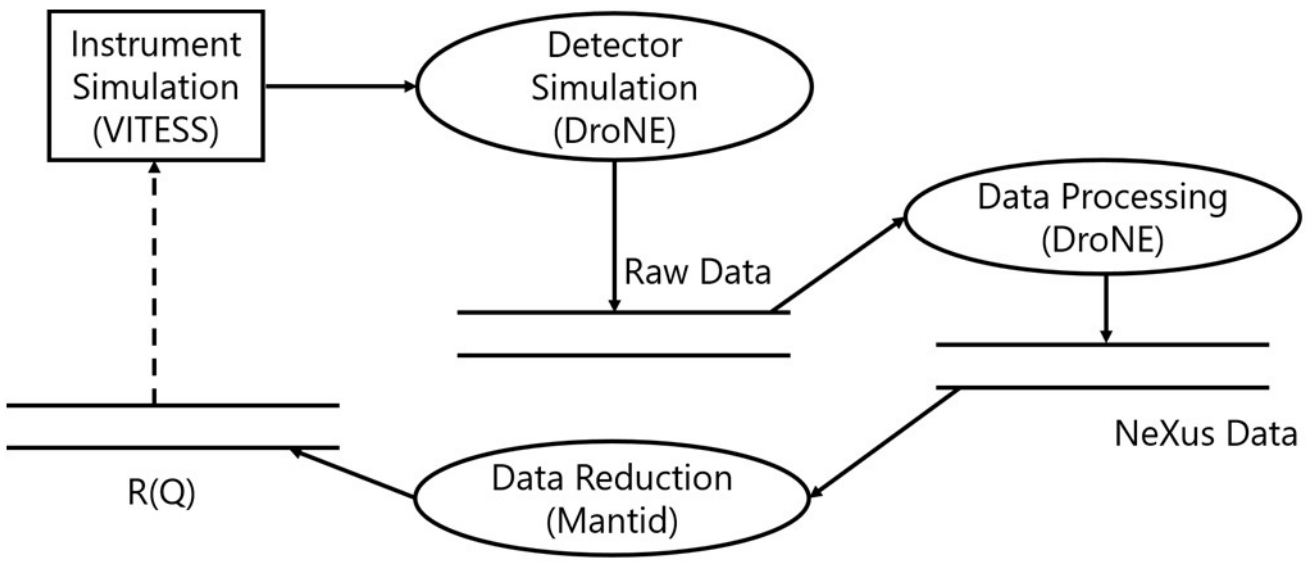

Fig.1 Data flow diagram of the virtual experiments

A Multi Wire Proportional Chamber (MWPC) $^3$He detector was simulated with DroNE by transforming the virtual neutrons trajectories from VITESS into electronics pulses. With detector geometry taken into consideration（2mm wire distance and 100% efficiency for all the neutrons）, the transformed electronic pulses was digitizes by DroNE, and then shored into raw data. The format of this raw data is the same as that used in the current developing instruments in the CSNS.

The data processing modules, which are also based on DroNE, reconstruct and map the received raw data and convert them into NeXus data files, where 2D histogram (intensity vs detector pixel id/TOF) are stored. Finally, data reduction based on Mantid was performed to covert the histogram data into two-dimensional data (reflectivity R(Q) versus scattering vector Q), where physical properties of the samples, like layer thickness and chemical composition, could be extracted from. Details of the instrument simulation, data processing and data reduction will be introduced as following.

## Instrument simulation

Based on the characteristic of the pulsed source of CSNS, a typical virtual neutron reflectometer with a moderate length of 21.5m was simulated as shown in Fig. 2. Main components of the instrument include one coupled hydrogen moderator source, neutron guides, choppers, slits, beam monitor, reflection sample and detector. The simulation started with bunch virtual neutrons at the surface of the moderator generated by sampling the two-dimensional intensity map of energy and time. The virtual neutrons were randomly distributed over the 10x10cm moderator surface with random angle relative to the beam direction. Every neutron was tagged with a weight factor, and sum of all the weight factors equals the total neutron flux (n/s/cm$^2$) at the moderator. Then the neutron was reflected several times by the inner surface of the straight neutron guides (inner size 40x40mm) and taper guides (converge from 40x40 to 20x30mm), a weight factor ranging from 0 to 1 was multiplied to the original weight factor according to the incident angle and reflectivity of the guides. Head and tail of the neutron waveband was cut by disc choppers, avoiding frame overlapping between neighboring pulses[28]. Incident beam at the sample is defined by two narrow slits. Reflected by the sample, the survived neutron will finally be collected by the detector. Generally, a bender (curved guides) or a T0 chopper are used in a reflectometer to avoid gamma rays emitted from the target. Benders and T0 chopper are not considered in this study to simplify the simulation and save consumptions of computer time.

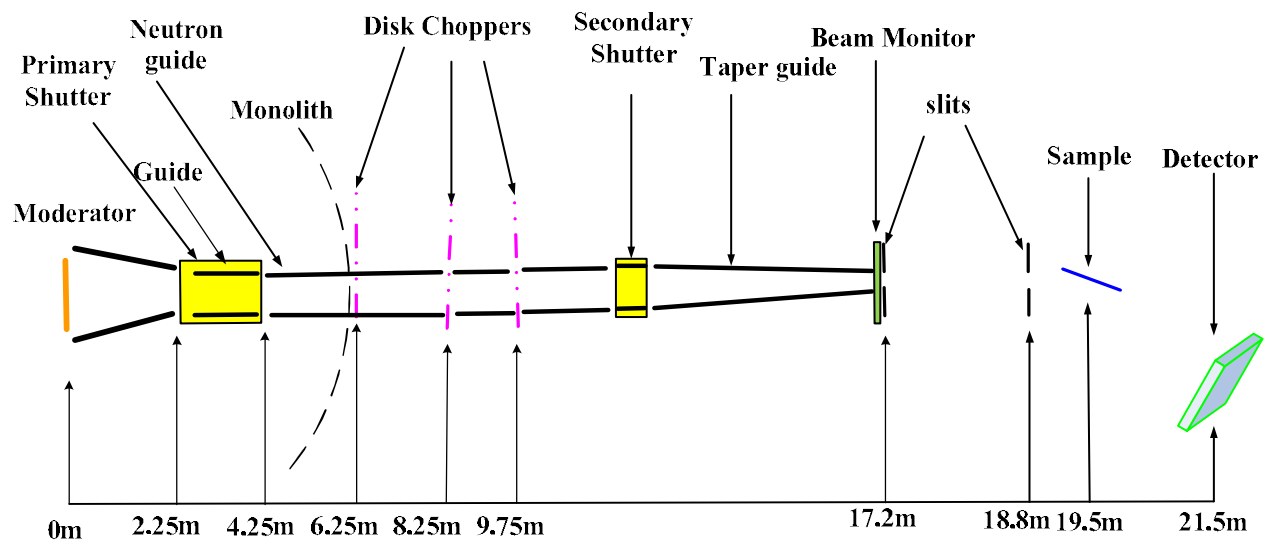

Fig. 2 Aerial view of the setup of the reflectometer

In a virtual reflection experiment as well as real reflection experiment, the range of momentum transfer Q should be determined first considering performance of the instrument as well as the properties of the sample. As for the specular reflection,

$$\begin{cases} Q_{min} = \frac{4\pi * \sin(\theta_{min})}{\lambda_{max}} \\ Q_{max} = \frac{4\pi * \sin(\theta_{max})}{\lambda_{min}} \\ \Delta Q \approx \frac{2\pi}{d} \end{cases} \quad (1)$$

Where, θ is incident angle, λ is wavelength of the incident beam and d means correlation length or layer thickness inside a specific sample. ΔQ should be at least within the range of $Q_{min}$ and $Q_{max}$.

It's clear from formula (1) that with a fix incident angle $\theta=\theta_{min}=\theta_{max}$ a wide range of Q or d can be covered in a TOF instrument. Usually, the whole Q-range of the instrument can be covered by two or three overlapped settings. Pulsed source of CSNS have a relative slow repetition rate of 25Hz (T=0.04s). Taking a total length of L=21.5m, and avoiding cross-talk between pulses, wavelength span at one setting can be calculated to be

$$\Delta\lambda = \left(\frac{h}{m_n}\right) * \frac{T}{L} = \frac{3.956*10^{-7} m^2/s}{21.5m} * 0.04s = 7.36 * 10^{-10}\text{m} = 7.36\text{Å} \quad (2)$$

where, h is Plank constant and $m_n$ is the weight of a neutron.

Neutron spectrum of a couple hydrogen moderator usually peaks at 2 to 3 angstroms. Therefore, we choose a waveband of 2 to 2+7.36=9.36Å. Note that the waveband could be shifted to longer wavelength, i. e. 8 to 15.36 Å, by shifting the phase of the disc choppers to access lower Q. With three angle settings 0.3°,1°and 3.6°, a wide Q range from $Q_{min}$ =0.007 to $Q_{max}$ =0.39Å$^{-1}$ could be covered with reasonable overlap as shown in Table 1.

Table 1 Slit openings and corresponding Q range

| **Incident Angles/°** | **slit1/mm** | **slit2/mm** | **$Q_{min}$ 1/Å** | **$Q_{max}$ 1/Å** |
|---|---|---|---|---|
| **0.3** | 0.2 | 0.1 | 0.007 | 0.033 |
| **1** | 0.65 | 0.325 | 0.023 | 0.11 |
| **3.6** | 2.34 | 1.17 | 0.084 | 0.39 |

A typical complete reflectivity curve R(Q)

normally includes the total reflectivity at the lowest Q side and about $10^{-6}$ reflectivity at highest Q side. In order to access the lowest point of total reflectivity, width of the high precision slits could be set as 50 to 200 micrometers with 1 to 2 micrometers uncertainty. The narrowest width of the slits are usually 50 micrometers with 1 to 2 micrometers uncertainty to access the lowest point of total reflectivity. Due to the $Q^{-4}$ decay of reflectivity[29] over medium and high Q, any means that could increase the flux of incident neutron beam are pushed to their limit, like relaxing the resolution and enlarging the sample size. As for the samples with very flat surface and interface, like sputtered thin films, relatively small samples (typically less than 2x2cm) were prepared and over illuminated as shown in Fig. 3. Slit width are usually chosen to keep a constant angular resolution Δθ/θ at different settings. As for the samples with rough surface or interface like liquid or polymer, big samples (typically 5x5cm or bigger) were prepared to get higher incident neutron current and got $10^{-6}$ reflectivity within a reasonable beam time.

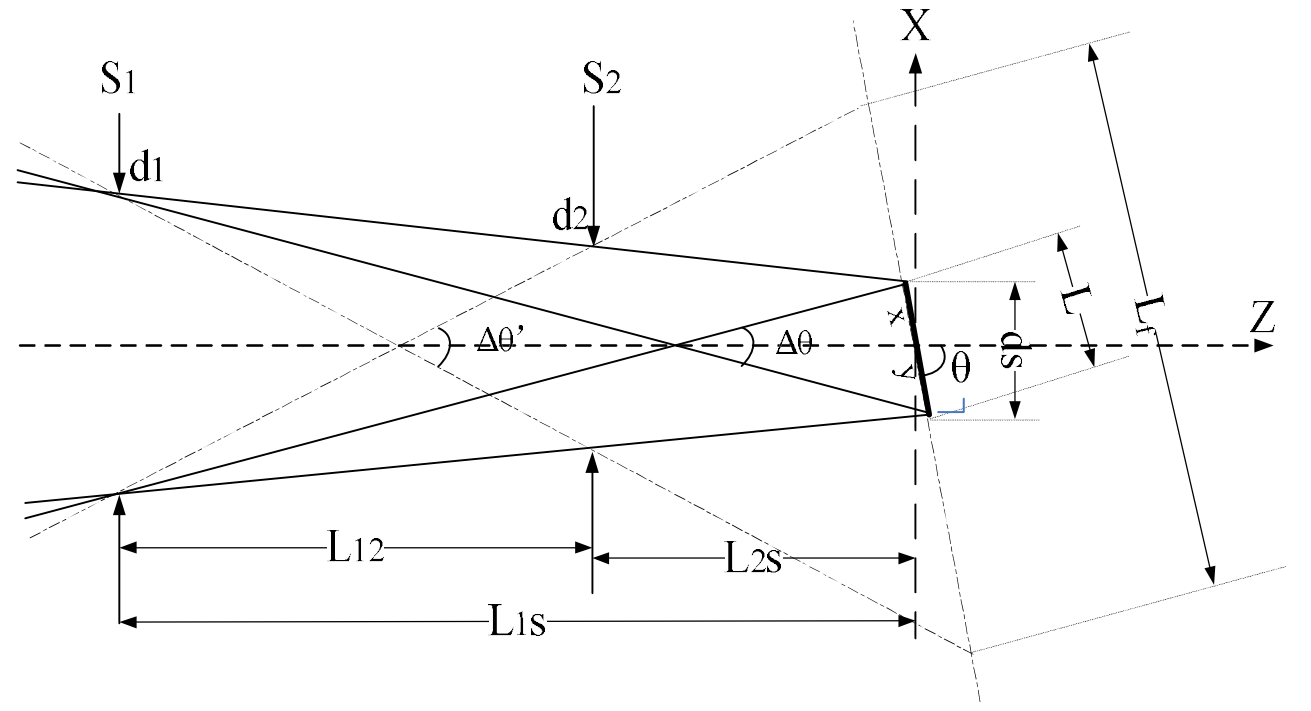

Fig. 3 Schematic view of the configuration of the slit system

A schematic view of the geometry of the slit system was plotted as shown in Fig 3. Coordinate was chosen in accordance with Fig 2, where Z is the incident beam direction, X means horizontal direction, Y points out of the paper. Slit1 ($S_1$) and Slit2 ($S_2$) are separated by $L_{12}$ and $L_{2s}$ from the sample. Imagine we have a sample with width L and required incident angle θ and angular resolution Δθ/θ. According to the geometry in Fig 3,

$$\begin{cases} x + y = L \\ ds + d1 = 2L_{1s} * \tan\left(\frac{\Delta\theta}{2}\right) \\ \frac{\frac{ds}{2}}{y*sin\theta} = \frac{\frac{dsL_{1s}}{ds+d1}}{ycos\theta+\frac{dsL_{1s}}{ds+d1}} \\ \frac{x*sin\theta}{\frac{ds}{2}} = \frac{-xcos\theta+\frac{dsL_{1s}}{ds+d1}}{\frac{dsL_{1s}}{ds+d1}} \end{cases} \tag{3}$$

Solve the equations of (3)

$$\begin{cases} d_s = L\left[\sin\theta - \mathrm{ctan}\theta\cos\theta\tan\left(\frac{\Delta\theta}{2}\right)^2\right] & (4) \\ d_1 = 2L_{1s} * \tan\left(\frac{\Delta\theta}{2}\right) - d_s & (5) \\ x = \frac{1}{\left[\frac{2sin\theta}{d_s} + \frac{d_s+d_1}{d_sL_{1s}}cos\theta\right]} & (6) \\ y = \frac{1}{\left[\frac{2sin\theta}{d_s} - \frac{d_s+d_1}{d_sL_{1s}}cos\theta\right]} & (7) \end{cases}$$

Then $d_2$ could be derived as

$$d_2 = d_s + (d_1 - d_s) * \frac{L_{2s}}{L_{1s}} \tag{8}$$

As shown in Fig 3, if the width of $S_1$ and $S_2$ was fixed, and width of the sample L was extended up to footprint of incident beam $L_f$,

$$L_f = \left[d_2 + \frac{L_{2s}}{L_{12}} * (d_1 + d_2)\right] * \frac{1}{\sin(\theta)} \tag{9}$$

divergence of the incident beam would be determined by $S_1$ and $S_2$ as Δθ' and corresponding angular resolution Δθ'/θ was defined by

$$\frac{\Delta\theta'}{\theta} = \frac{2\,\mathrm{atan}\left(\frac{d_1-d_2}{2L_{12}}\right)}{\theta} \tag{10}$$

In the virtual experiment of this work, a constant angular resolution Δθ/θ=0.021 and a sample width L=11.5mm was chosen. Slit width parameters in Table 1 are calculated by substituting all the parameters into formula (4) (5) and (8) (with $L_{12}$=1680mm and $L_{2s}$=635mm). Then sample width L was enlarged to $L_f$ with fixed $d_1$ and $d_2$ to make sure that all the incident neutrons are reflected. Whereas, angular resolution would be degraded to be $\Delta\theta'/\theta=0.033$ as calculated with (6), which is still acceptable in our case. Vertical openings of the slits are usually set to be the same as sample height which is 40mm here.

**Sample simulation**

Reflectometer measures the Scattering Length Density (SLD) differences between layers of thin films via the interference of neutron wave between layers of thin films. However, in a virtual experiment, neutrons are treated as trajectories or particles with certain speed or wavelength, the duality of neutrons and the quantum effect of neutron wave interference are out of the range of Monte-Carlo simulation. When a virtual neutron with wavelength $\lambda_i$ reached the surface of the sample with incident angle $\theta_i$, its momentum transfer Q was calculated by

$$Q_i=4*\pi\sin\theta_i/\lambda_i \qquad (11)$$

where i means $i^{th}$ neutron. Weight factor of the neutron was multiplied by a value according to the pre-calculated reflectivity curve $R_0(Q)$ (Parratt calculated curve in Fig 4). Since all the geometry effect like incident beam divergence and detector resolution are taking into consideration, $R_0(Q)$ would be smeared just like the real experiments.

Thick layers usually cause fine fringes in low-Q part of the reflectivity curve, on the contrary, thin layers cause gentle oscillations over a wide range of Q. Bearing this in mind, we choose a prototype sample Si(substrate)/Fe(200Å)/$V_2O_5$(1800Å) which means a thin layer of Iron(200Å) sputtered on silicon substrate and then covered by a thick layer of Vanadium (1800Å). With such sample, both the smearing effect of thick layer oscillations and contour of the thin layer oscillations can be demonstrated. Input file of the sample component was calculated with Parratt32 software without considering the roughness of the surface and interface.

As shown in the Fig 1 and Table 1, with three angles and corresponding slit width, data were collected with (reflected beam) and without sample (direct beam) at the monitor and detector site. $I_{Ref}$(TOF) $I_{Ref_moni}$(TOF) and $I_{Direct}$(TOF), $I_{Direct_moni}$(TOF) of each angle was extracted and transferred to $I_{Ref}(\lambda)$, $I_{Ref_moni}(\lambda)$ and $I_{Direct}(\lambda)$, $I_{Direct_moni}(\lambda)$ with the relation of formula (2). $R(\lambda)$ was obtained with

$$R(\lambda) = \frac{I_{Ref}(\lambda)}{I_{Ref_moni}(\lambda)} \Big/ \frac{I_{Direct}(\lambda)}{I_{Direct_moni}(\lambda)} = \frac{I_{Ref}(\lambda)}{I_{Direct}(\lambda)} \qquad (12)$$

Then R(Q) was obtained by transferred $R(\lambda)$ to R(Q) with

$$Q_j=4*\pi\sin\theta/\lambda_j \qquad (13)$$

where j means $j^{th}$ $\lambda$ bin and $\theta$ is the incident angle. R(Q) from the simulation results and Parratt formula are compared as shown in Fig. 4. Fine fringes of V-layer can be clearly observed over the whole Q-range of the calculated curve, while fringes higher than $Q=0.2Å^{-1}$ are totally smeared out in the simulated curve. Such smearing can be explained that at about $Q=0.2Å^{-1}$ oscillation width of the V-layer

$$\frac{2\pi}{d_{V_layer}} = \frac{2\pi}{1800} = 3.49 \times 10^{-3}Å^{-1}$$

almost equals the uncertainty of momentum transfer

$$Q\left(\frac{\Delta Q'}{Q}\right) = \frac{0.2\left(\frac{\Delta\theta'}{\theta}\right)}{2} = 3.3 \times 10^{-3}Å^{-1}$$

While, oscillations of the iron layer fit quit well and the smearing effect is not so obvious.

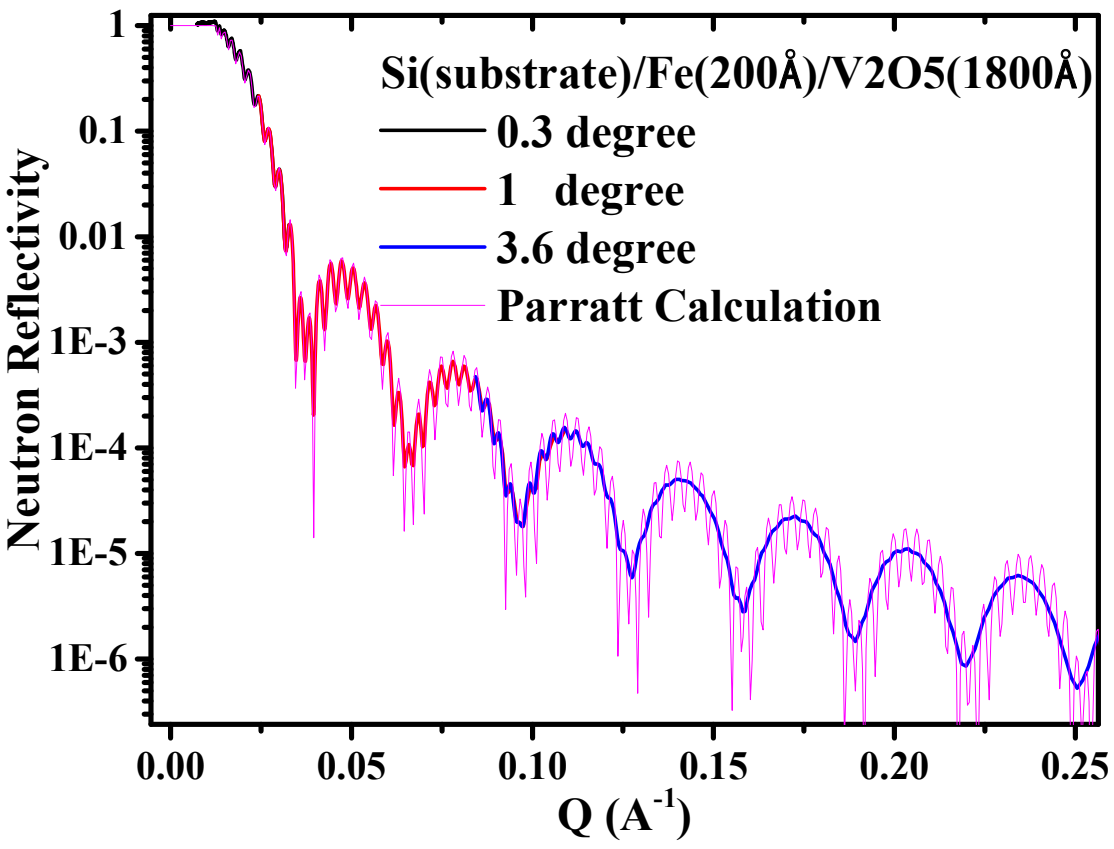

Fig. 4 Reflectivity curve of the sample compared with Parratt calculation

In a real experiment, live data at the detector can be plotted by integrating coordinate at the detector x, y or λ to show users quality of the data and do some primary data analysis. As shown in Fig 5 Lambda-x distribution I(x, λ)(n/s/cm/Å) of the reflected beam of the three angles and I(λ) of the incident and reflected beam are plotted. Fringes and oscillations can be clearly seen in the I(x, λ) plot. In the I(λ) plot, total reflection at 0.3 deg. can be identified from 5 to 9Å.

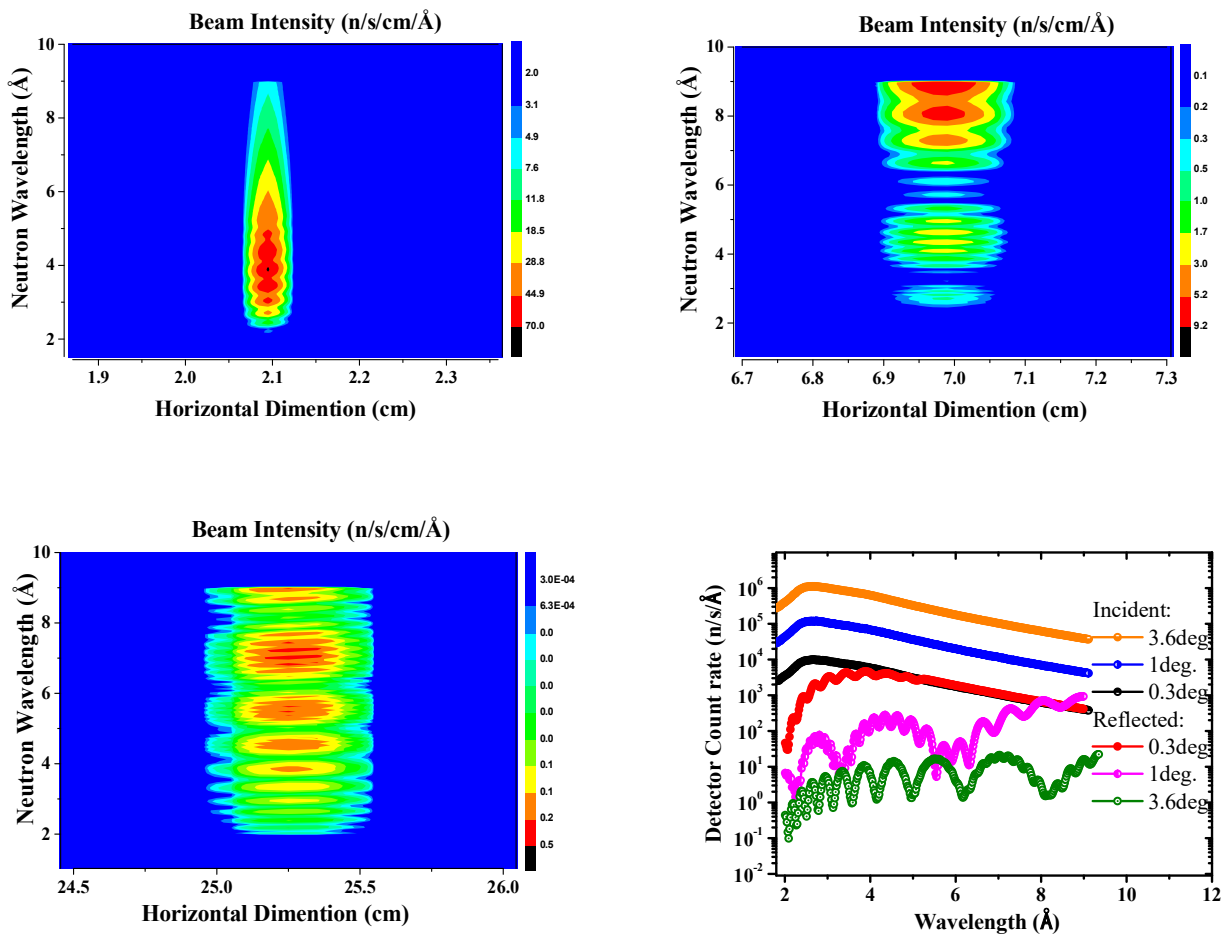

Fig. 5 Horizontal intensity distribution at the detector Vs. neutron wavelength or I(x, λ) plot with incident angle of 0.3 degree (up left), 1degree (up right) and 3.6 degree (down left). And I(λ) plot of the incident and reflected beam (down right).

I(λ) of Fig 5 was integrated to get the total neutron current of incident and reflected beam as shown in Table 2. Reflected neutrons decrease dramatically with increasing angle and it takes hours to collect enough counts at 3.6 degree.

Table 2 Integrated neutron current of the incident and reflected beam

| Incident Angles/° | Incident Current (n/s) | Reflected Current (n/s) |
|---|---|---|
| **0.3** | 1.85E+04 | 9.30E+03 |
| **1** | 1.40E+05 | 6.50E+02 |
| **3.6** | 2.90E+06 | 36 |

**Detector simulation**

Simulated data can be analyzed directly as shown in the sample simulation section. However, the trajectories exported is far from the raw data of real experiments. Detector simulations are done to convert the data into electronic pulses and raw data for data reduction. Three sets of data are exported. First, reflected virtual trajectories to mimic the events data collected by the detector; second, transmitted virtual trajectories to simulate the direct beam; third, λ binned intensity collected by the monitor. It must be noted that a virtual trajectory is not a neutron event. It is recorded with a precise position, lambda, TOF, spin and a weight factor. But weight factor doesn't exist since we could never detect a fraction of a neutron. Therefore, all the weight factors of virtual neutron trajectories were multiplied by a certain number and round off. Then the weight factors were big integers and were treated like neutron events at the detector.

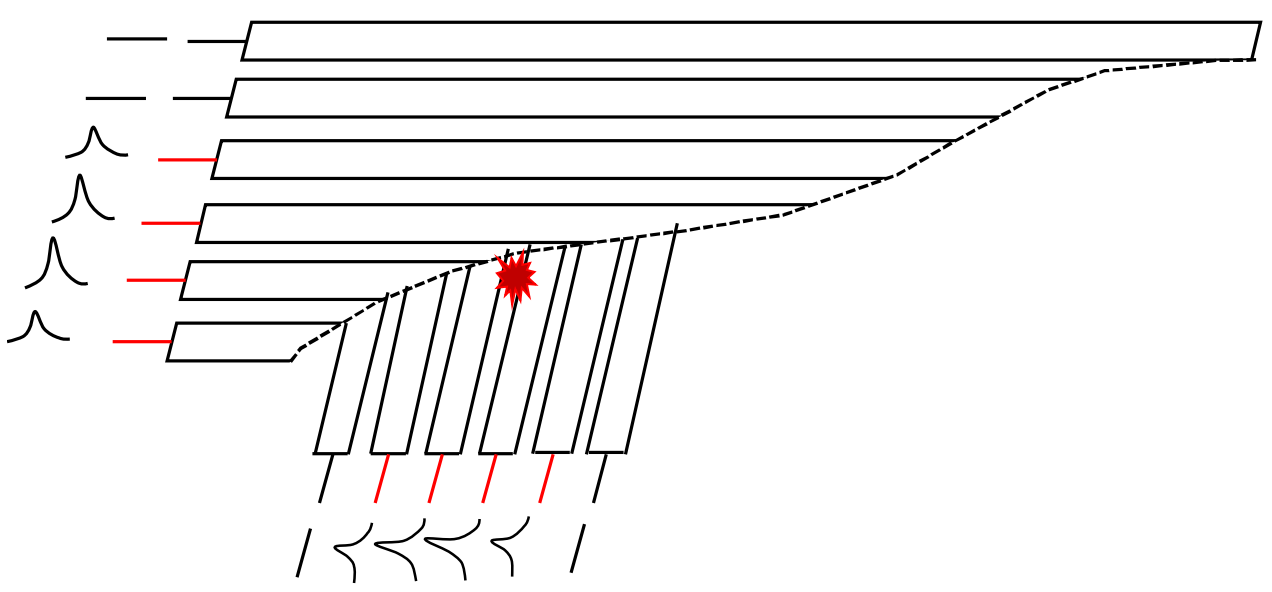

Fig. 6 Crossed wires of the MWPC Position Sensitive Detector and charge pulses fired by one neutron-hit event.

All the neutron events are exposed to the detector simulation model to fire the readout strips and generate the output signals. A typical Multi-Wire Proportional Chamber (MWPC) Position Sensitive Detector (PSD) is involved in this study. The size of detector is 160x160mm with horizontal and vertical readout plants to pin point the position of the hit event. As shown in Fig 6, each readout plant consists of 40 3mm wide readout strips separated by 4mm. The principle and detailed structure of the neutron detector can be found in Ref. [29,30].

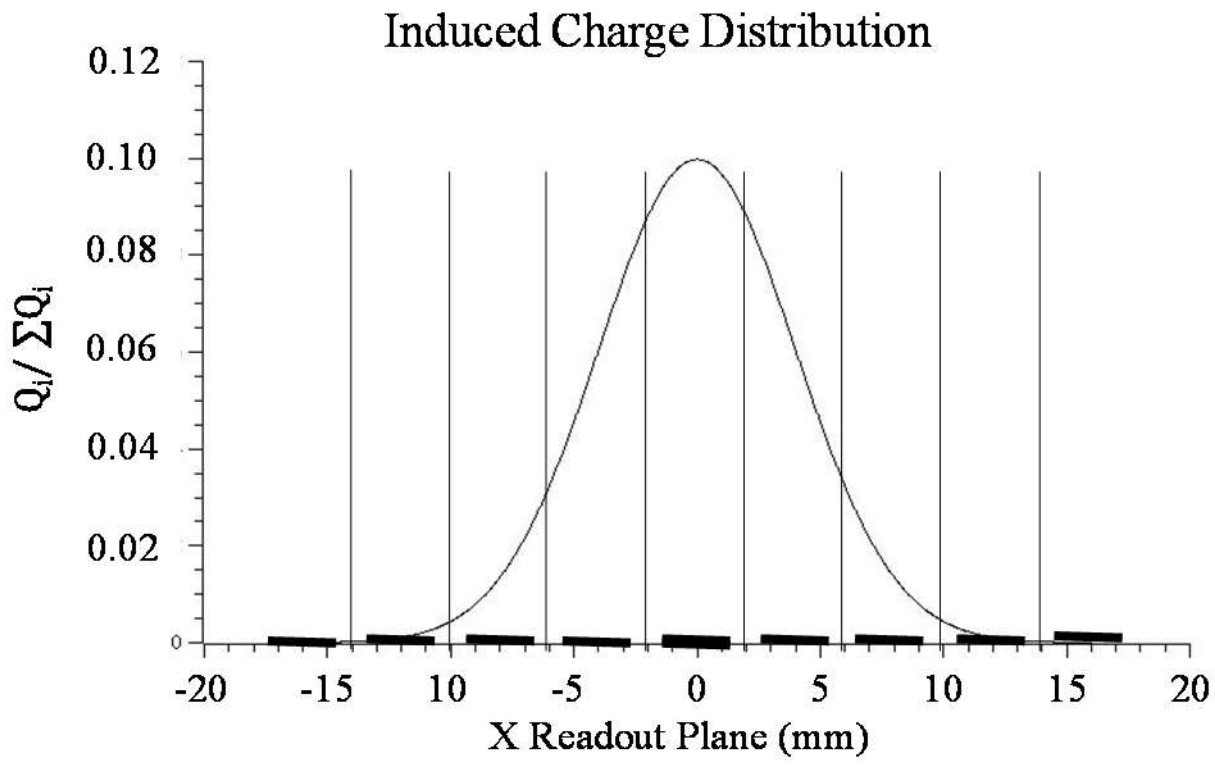

Fig. 7 The induced charge distribution on the X readout plane; X axis represents the position on the readout plane; Y axis is the ratio of induced charge to the total charge. The charge between two vertical lines is collected by the center readout plane indicated as thick black lines.

When a neutron traverses the detector, its time of flight (t) is recorded and the induced charge on the readout strips of MWPC are generated according to the Gauss distribution as shown in Fig 7. With standard deviation of 4mm, at least 3 strips on each readout plant can be fired. Thereby, one neutron event will generate at least 6 records in the raw data. Each record contains the charge collected by the strip and the arriving time of the signal.

Pulse-level information needs to be defined in the simulation configuration, such as the pulse time and electronics module id. All these data are packed as raw data and exported for the reconstruction algorithms.

## Reconstruction of the raw data

Mass data of electronic pulses from the detector simulation need to be reconstructed to the events data. A neutron event is described by its position and TOF. TOF of the event is given by the timing of the first arriving signal and the position P(X,Y) is calculated according to the center-of-gravity method. The charge of induced pulses on the readout plant are stored in the raw data, and the center of gravity of the charge on the horizontal strips is computed according to the formula

$$P(X) = \frac{\Sigma Q_i X_i}{\Sigma Q_i} \quad (14)$$

Where $Q_i$ is the charge collected by the $i^{th}$ strip, and $X_i$ is the position of the strip. Center of gravity on the vertical strips can be calculated in the same way. With P(X, Y) from the center-of-gravity method, a mapping algorithm is used to determinate the pixel ID. The size of pixel in this work is 2x2mm, and there are 6400 pixels in total. Finally, the 2D histogram (Intensity vs. Pixel ID/TOF) are saved as NeXus files.

## Reduction of the data

The aim of the data reduction algorithm is to obtain a complete reflectivity curve R(Q) from the NeXus files. A brief illustration of the workflow of the reduction algorithm was shown in Fig. 8.

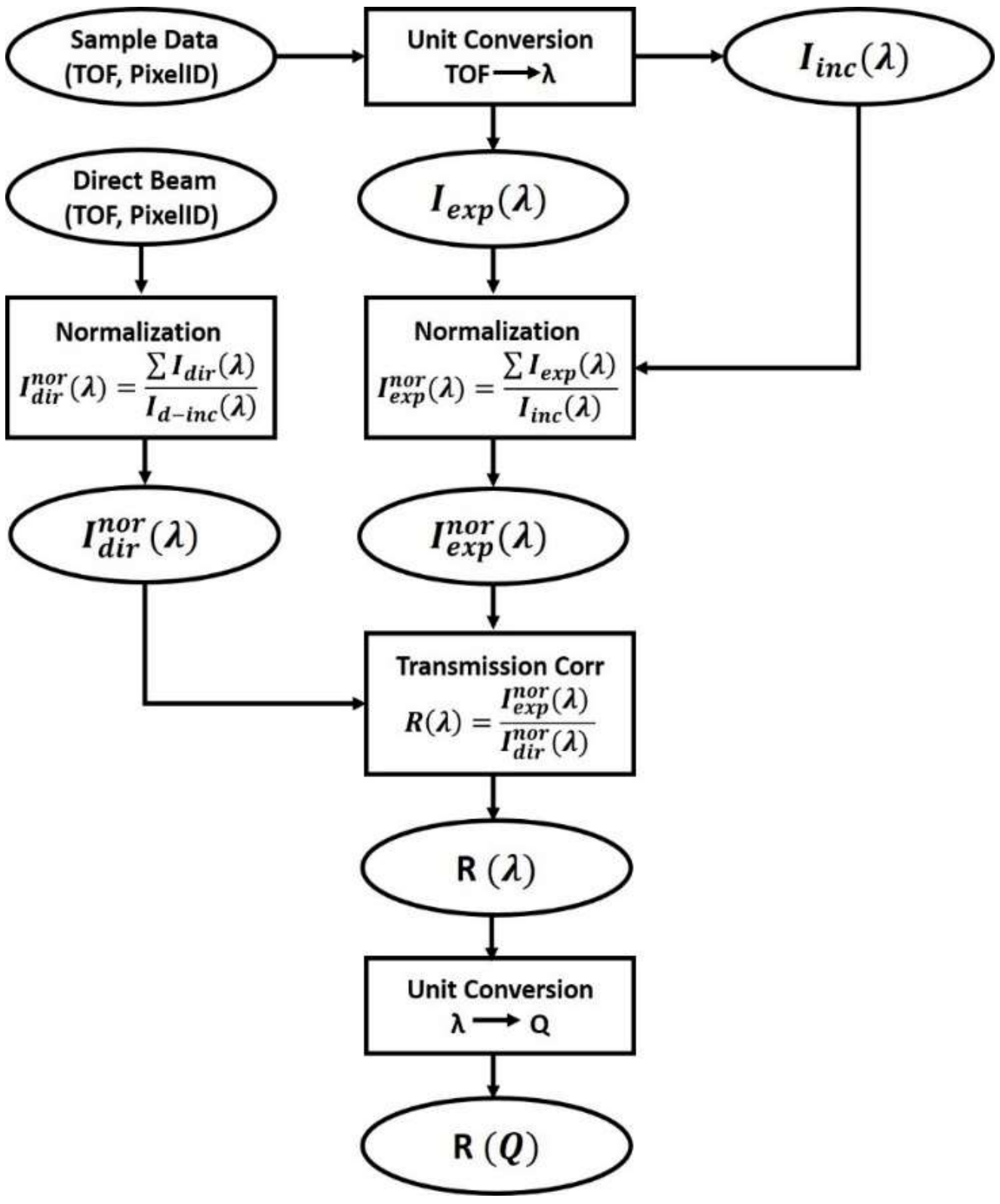

Fig. 8 Overview of the data reduction workflow for TOF reflectometry

As shown in Fig. 8, two sets of data are imported from data reconstruction, one is reflected data and the other is direct beam data. Both data are normalized by its own incident flux measured by the monitor. Before the normalization, both data sets need to be converted from TOF to wavelength according to

$$\lambda = \frac{h \cdot T}{m \cdot L} \qquad (15)$$

where T represents the TOF stored in the NeXus files, h means Planck constant, and m is the mass of neutron, L is the flight path from the neutron source to a given pixel of the detector or monitor. Ideally, if the neutron source is steady and provide the same neutrons at certain time span (which is the case in the virtual instrument simulation), the reflectivity curve can be obtained by just divide the reflected curve with the direct beam. However, neutron flux at the sample point may fluctuate over time. Thereby, both the reflected beam $\sum I_{ref}(\lambda)$ and direct beam $I_{dir}(\lambda)$ have to be normalized by their own incident beam recorded by the monitor before the sample. The reflectivity R(λ) was calculated with the following formula

$$R(\lambda) = \frac{I_{ref}^{nor}(\lambda)}{I_{dir}^{nor}(\lambda)} \qquad (16)$$

Finally, R(Q) was obtained by transferring R(λ) to R(Q) according to the formula

$$R(Q_i) = R(\lambda_i) \qquad (17)$$

Where, $Q_i = \frac{4\pi}{\lambda_i}\sin(\theta)$

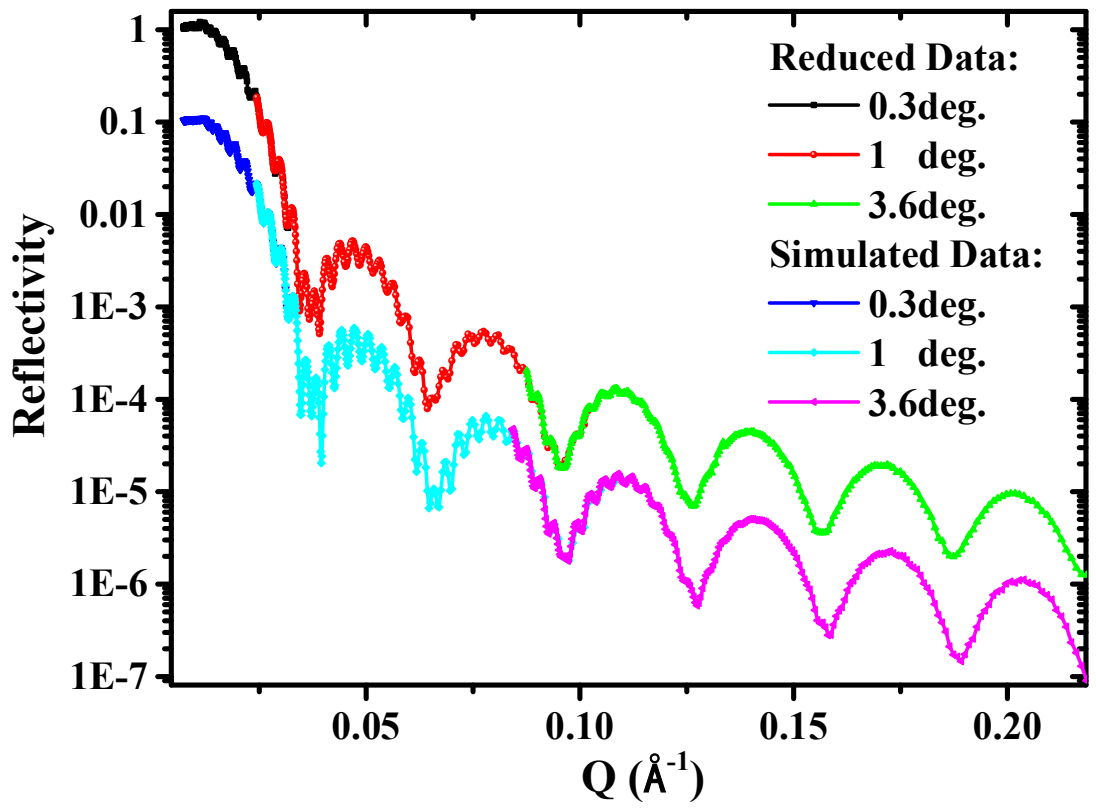

Fig.9 Neutron reflectivity curve of the reduced data and simulated data (the simulated data was shifted an order of magnitude downwards for

clarity)

As shown in Fig 9, reduced R(Q) was compared with the simulated curve from Fig 4. It fit quite well except a little smearing of the oscillations of the thick layer of Valium. The small smearing was caused by the resolution effect of the detector.

## Conclusion

Reliable data is the foundation that good science to be delivered in the coming instruments of CSNS. Based on a complete virtual experiment of a general TOF neutron reflectometer, a batch of data processing and data reduction tools and methods are established. Details of the instrument simulation and data reduction are also demonstrated.

## Acknowledgement

We are grateful to Dr. Klaus Lieutenant (HZB) for his help using VITESS, and Dr. Frank Klose for profitable discussion about the design of neutron reflectometer.